\documentclass[twocolumn,showpacs,preprintnumbers,prb]{revtex4}
\usepackage{graphics}
\usepackage{graphicx}
\newcommand{\res}{6$\tilde{\sigma}$}
\begin{document}
\title{Adsorption-induced constraint on delocalization of electron states in an Au chain on NiAl(110)}
\author{Mats Persson}
\email{tfymp@fy.chalmers.se} \affiliation{ Department of Applied
Physics, Chalmers/G\"oteborgs University, S-41296, G\"oteborg,
Sweden}
\date{\today}

\begin{abstract}
We have carried out a density functional study of the localized
constraint on the delocalized, unoccupied resonance states in a
mono atomic Au chain on a NiAl(110) surface by adsorption of a CO
molecule on one of the adatoms in the chain. This constraint was
observed recently by scanning tunnelling microscopy and
spectroscopy. The repulsive interaction of the occupied
5${\sigma}$ molecular state with the unoccupied, resonance state
in a Au adatom in the chain is found to break the degeneracy of
the Au adatom-induced resonance states and to disrupt their
delocalization in the chain.
\end{abstract}
\pacs{73.20.Hb, 73.21.Hb, 68.37.Ef} \maketitle

Adsorbate-induced modifications of confined electronic states in
metallic nanostructures on surfaces such as supported atomic wires
are of direct interest in chemical sensing and also in the
emerging field of molecular electronics through their influence on
electron
transport~\cite{AtWireRev,Konetal,Coletal,Favetal,Rometal}. To
gain physical insight about these complex phenomena, one needs to
study how the confined electronic states of a single, adatom
nanostructure with well-defined atomic structure and composition
is affected by controlled adsorption of a single atom or molecule.
This goal was recently realized by a scanning tunnelling
spectroscopy study of fabricated monoatomic Au atom chains of
various lengths on a NiAl(110) surface upon adsorption of a single
CO molecule at various sites along the chain~\cite{NilWalHo03}.
The bare chains have been shown to support "particle-in-box"-like
unoccupied electron states~\cite{NilWalHo02}. Spatially resolved
measurements of these electronic states revealed that the adsorbed
CO molecule introduced a localized constraint on the electron
delocalization in the chain. Thus there is a need to understand
the physics behind this adsorption-induced constraint from a
theoretical study of these electronic states.

The electronic structure of monoatomic Au adatom chains on a
NiAl(110) surface has been studied recently by density functional
calculations~\cite{CalCavNar04,Per04}. In particular, the study in
Ref.~\onlinecite{Per04} have revealed the nature of the
unoccupied, resonant states in these chains and their
"particle-in-box" character. The observed energy splittings of the
resonance doublets of Au addimers as a function of the interatomic
distance~\cite{NilWalPerHo03} and the observed energy dispersion
of the resonance states in Au chains were well reproduced by the
calculated, unoccupied Kohn-Sham
states~\cite{footnoteExcitedState} in this study. The resonance
states were shown to derive from the 6$s$ states of the Au atoms,
which develop a $p_z$ character upon adsorption. The resonance
states were found to interact strongly with the substrate states
as reflected by long-range, substrate-mediated interactions
between adatom resonances. In contrast, Calzolari and
coworkers~\cite{CalCavNar04} argued from their density functional
study of transport in Au ad-chains that they behaved as isolated
Au chains. They also studied the electronic and geometric
structure of adsorption of a CO molecule on a seven Au adatom long
chain. A repulsive interaction induced by the Au-C bond was found
to induce a interaction that pushed the Au 6$s$ level to higher
energies. Because they did not identify any unoccupied resonance
states of the adsorbed Au chain, it is not clear how this
repulsion results in a localized constraint on the delocalization
of these states in the adsorbed chain.

In this paper, we report a density functional study of the
influence of an adsorbed CO molecule on the electronic states of
an Au adatom and adtrimer on a NiAl(110) surface. These adatom
structures albeit being of minimal size reveal directly the
physics behind the localized constraint on the delocalization of
the unoccupied electron states. This physics can be directly
generalized to Au adatom chains of arbitrary lengths.

In order to reveal the electronic and geometric effects of an
adsorbed CO molecule on mono-atomic Au chains on NiAl(110), we
have carried out density functional calculations of CO molecules
adsorbed on an Au adatom and on an edge and center adatom of an Au
ad-trimer. As in the previous study of the Au adatom and
ad-trimer~\cite{Per04}, these calculations were carried out using
the projector augmented wave method as implemented in the {\tt
VASP} code~\cite{VASPrefs}. The exchange and correlation effects
were represented by the PW91 version~\cite{PW91} of the
generalized gradient approximation. The systems were represented
by a CO molecule and Au atoms on a NiAl slab in a super cell
geometry~\cite{cell} and the equilibrium structures were obtained
by structural optimization\cite{geometry}. The nature of the
electronic states was investigated by calculating the partial wave
decomposition of the density of states within atomic
spheres~\cite{PDOS} and the local density of states~\cite{LDOS}
outside the surface.

\begin{figure}
\includegraphics[width=6cm, angle = -90]{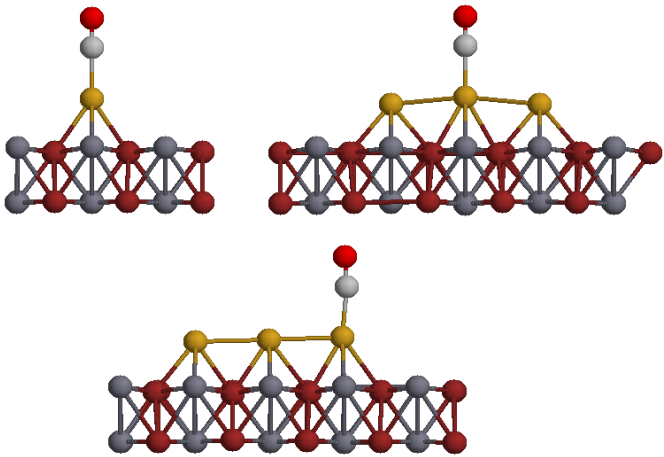}
\caption{Ball and stick model of the calculated geometric
structure of an CO molecule adsorbed on (A) a single Au adatom,
and (B) the center adatom and (C) the edge adatom of an Au
ad-trimer. Only the two outermost layers of the NiAl(110)
substrate are shown. \label{fig:geom}}
\end{figure}

The calculated equilibrium structures of a CO molecule on a single
Au adatom and Au ad-trimer are depicted in Fig.~\ref{fig:geom}.
The equilibrium structures of the bare single Au adatom and
ad-trimer were already discussed in an earlier density functional
study~\cite{Per04} and are not discussed here. The CO molecule is
adsorbed in an upright position on a single Au adatom and on the
center atom of the Au ad-trimer and in a slightly tilted
configuration on an edge atom of the Au ad-trimer~\cite{COgeom}.
The adsorption of the CO molecule on the edge atom is only about
0.023 eV more stable than the adsorption site on the center atom.
This finding is consistent with the experimental observation that
tunnelling at high bias and current tended tended to displace the
CO admolecule from the interior to the ends of the
chains.~\cite{NilWalHo03}. The main effect of the CO adsorption on
the Au adatoms is an upward shift of the Au adatom coordinated to
the O atom by about 0.1 {\AA} for the center adsorption site on
the Au ad-trimer and by about 0.2 {\AA} for the other two
structures.

\begin{figure}
\includegraphics[width=6cm, angle = -90]{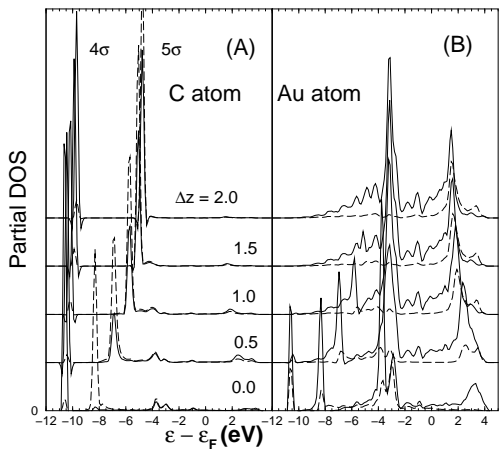}
\caption{Evolution of calculated partial density of $s$ (solid
lines) and $p_z$ (dashed lines) states within a sphere around (A)
a C atom and (B) a Au atom in CO/Au/NiAl(110) as a function of
distance $\delta z$ of the CO molecule from its equilibrium
position on the Au adatom. The peaks in the partial DOS for the C
atom derive from the $4\sigma$ and $5\sigma$ molecular orbitals of
the CO molecule as indicated. \label{fig:EvolPartDOS}}
\end{figure}

The effect of the adsorption of a CO molecule on a single Au
adatom on its unoccupied, resonance state {\res} is revealed by
studying the evolution of the partial DOS around the Au adatom
(Fig.~\ref{fig:EvolPartDOS}A) and the C atom
(Fig.~\ref{fig:EvolPartDOS}B) as a function of the CO molecule
distance, $\delta z$, from the Au adatom. At $\delta z = 2$ {\AA},
the electron states of the CO molecule and the Au adatom are
essentially decoupled. The {\res} state of the Au adatom show up
as a relatively narrow peak at about 1.6 eV above the Fermi level
in the DOS of $s$ and $p_z$ partial waves around the Au adatom.
The partial DOS with $s$ character covers a wide energy range and
exhibits also a relatively narrow resonance structure at 3.5 eV
below the Fermi level. The DOS for $s$ and $p_z$ partial waves
around the C atom are dominated by the 4$\sigma$ and 5$\sigma$
molecular orbitals of the CO molecule. Around the C atom,
4$\sigma$ has predominantly $s$ character, whereas 5$\sigma$ has
predominantly $p_z$ character. The 1$\pi$ and the 2$\pi$ molecular
orbitals of the CO molecule are symmetry forbidden to couple with
the {\res} state and are not shown in Fig.\ref{fig:EvolPartDOS}A.

At smaller $\delta z$, the 5$\sigma$ state starts to interact with
the {\res} state, resulting in a repulsion between these two
states and an interchange of their characters. The {\res} state
shifts up with about 1.5 eV and the 5$\sigma$ state shifts down
with about 3 eV. In contrast, the 4$\sigma$ state shows only a
minor downward shift in energy. Note that the strong interaction
between the 5$\sigma$ state and the {\res} state is manifested by
the repulsion of about 4.5 eV between these two states being
substantial in comparison between their original energy separation
of about 6 eV. The 5$\sigma$ state keeps its $p_z$ character on
the C atom but loses its $s$ character and develops an admixture
of predominantly $s$ character on the Au adatom. In contrast, the
{\res} state keeps its $s$ character on the Au adatom but loses
its $p_z$ character. The resonance structure around -3.5 eV in the
$s$ DOS around the Au adatom is weakly affected by the interaction
with the CO molecule.

In the Au adchains, the degenerate {\res} states of the individual
adatoms interact and form delocalized resonant states. The upward
energy shift of the {\res} state of an Au adatom by the CO
adsorption on the Au chains breaks its degeneracy with the other
{\res} states in the chain and has important consequences on the
delocalization of the {\res} states in the chain. These
consequences are here illustrated for CO adsorption on the Au
trimer on the NiAl(110) surface. First we need to describe the
nature of resonance states of the bare adtrimer.

\begin{figure}
\includegraphics[width=6cm, angle = -90]{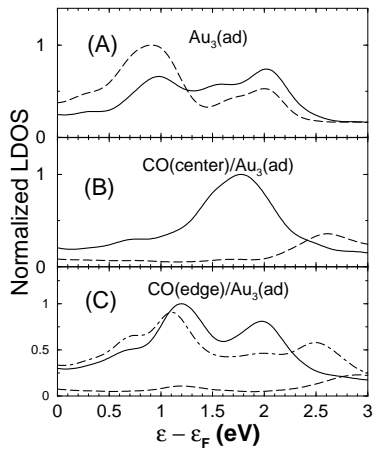}
\caption{Normalized local density of states (LDOS) for (A) Au
ad-trimer and a CO molecule adsorbed on (B) the center adatom or
(C) the edge adatom of the Au adtrimer. The LDOS has been
calculated at a distance of about {1 \AA} above an edge adatom
(solid and dashed-dotted lines) and the center (dashed lines)
adatom in the Au ad-trimer. In (C), the two edge atoms are
inequivalent and the solid and the dashed-dotted line gives LDOS
above the edge atom without and with an adsorbed CO molecule
,respectively. The LDOSs are normalized to their maximum values
above the center adatom in (A) and (C), and the edge adatom in
(B). \label{fig:TriLDOS}}
\end{figure}

The interactions among the three degenerate adatom {\res} states
in the trimer results in a resonance triplet with large energy
splittings as revealed by the calculated LDOS on-top of the center
and edge adatoms in Fig.~\ref{fig:TriLDOS}A. The first and the
third state at 1.0 and 2.1 eV, respectively, are both symmetric
where the former state has the largest amplitude on the center
adatom and the latter state has its largest amplitude on the edge
adatoms. The second state at 1.6 eV is antisymmetric and has
accordingly no amplitude on the center adatom. The interaction
between the {\res} states of the two edge adatoms is weak so that
the energy of this state is close to the energy of 1.7 eV for the
{\res} state in a single adatom. The energies of the first and
third state are symmetrically split around this energy.

Upon adsorption of a CO molecule on an adatom in the adtrimer, the
resonance structure of the adtrimer changes dramatically
(Fig.~\ref{fig:TriLDOS}). For the center adatom adsorption site,
the resonance triplet collapses to a resonance singlet
(Fig.~\ref{fig:TriLDOS}B), whereas for the edge adatom it
collapses to a resonance doublet (Fig.~\ref{fig:TriLDOS}C). The
resonance singlet and doublet energies are very close to the
values 1.7 eV for the {\res} state in a single adatom and 1.2 and
2.15 eV for the resonance doublet in the Au addimer, respectively.
The adsorption-induced breaking of the degeneracy of the {\res}
state in the adatom coordinated to the CO molecule with the {\res}
states of the remaining adatoms effectively turns off its
interaction with these latter states. In the case of adsorption on
the center adatom, the interaction between the {\res} states of
the two edge atoms, as mediated in the bare adtrimer essentially
by the {\res} state of the center adatom, is turned off. This
results in essentially two degenerate and isolated resonance
states. In the case of CO adsorption on an edge adatom, there is
still a direct interaction between the two {\res} states on the
remaining two adatoms, which splits them into a doublet.

The calculated changes in the resonance energy structure of the
adtrimer upon adsorption of a CO molecule can be rationalized in a
simple, tight binding resonance model. This model was introduced
in an earlier paper~\cite{Per04} and applied to various linear Au
adatom structures. The resonance triplet energies and states of
the bare adtrimer can be modelled by a value of 1.6 eV for the
on-site energy $\epsilon_0$ for the adatoms and a value of -0.35
eV for the nearest-neighboring off-site energy $t$. Note that
values of $\epsilon_0$ and 2$|t|$ are close to the values for the
calculated energy 1.7 eV of the {\res} state in the adatom and the
calculated energy splitting 0.9 eV of the resonance double in an
addimer. In this model, the resulting triplet energies are given
by $\epsilon_0+ t\sqrt{2}$ = 2.1 , $\epsilon_0$ = 1.6, and
$\epsilon_0- t\sqrt{2}$ = 1.1 eV and are close the calculated
resonance energies of 2.1, 1.6 and 1.0 eV. As suggested by the
calculated upward energy shift of about 1.5 eV for the {\res} by
CO adsorption on a single adatom, the effect of the corresponding
adsorption of a CO molecule on an adatom in the adtrimer is then
modelled by increasing the associated on-site energy by
$\Delta\epsilon$ = 1.5 eV. For simplicity, the off-site energies
are assumed to be unaffected by the adsorption. The
adsorption-induced effects on the electronic states of the
ad-trimer is understood semi-quantitatively in this model by
perturbation theory in the relatively small parameter
$|t|/\Delta\epsilon$.

We apply this model first to the case of CO adsorption on the
center adatom. To zeroth order in $t$, only the energy of the
resonance state of the center adatom is shifted up by
$\Delta\epsilon$. The degeneracy of the two resonance states of
the edge atoms is split first to second order in $t$. To this
order these states are split by $2|t|^2/\Delta\epsilon = 0.16$ eV
and both shift down by $2|t|^2/\Delta\epsilon = 0.16$ eV. Thus the
energy of antisymmetric state in this resonance doublet is
unperturbed and is the same as for the anti-symmetric state of the
bare ad-trimer. This behavior reflects that the perturbation by
adsorbed CO molecule is localized to the center adatom and does
not break the symmetry of the ad-trimer. The energy splitting of
the resonance doublet is small compared to the resonance width and
is not resolved in the calculated LDOS so the doublet appears as a
single broad resonance.

In contrast to CO adsorption on a center adatom the adsorption on
an edge atom breaks the symmetry of the ad-trimer. The resonance
energy of this edge atom is shifted up by $\Delta\epsilon$ and the
degeneracy of the two resonance states at the center adatom and
the bare edge adatom is split to leading order in $t$ by $2|t|$.
The resulting resonance doublet has then the same energies as for
the resonance doublet of the ad-dimer. The corrections to these
energies by second order perturbations in $t$ are small. The
resonance doublet energies shift down only by about
$|t|^2/(2\Delta\epsilon) = 0.04$ eV and the effect on the energy
splitting is of higher order in $t$.

In summary, we have carried out a density functional study of the
effects on the delocalized, unoccupied resonance states in a mono
atomic Au chain on a NiAl(110) surface upon adsorption of a CO
molecule on one of the adatoms in the chain. This study explains
the observed localized constraint by the adsorbed CO molecule on
these states in terms of a repulsive interaction of the occupied
5${\sigma}$ molecular state with the unoccupied, resonance state
in the Au adatom coordinated to the adsorbed CO molecule. This
interaction breaks the degeneracy of the Au adatom-induced
resonance states in the chain and constrains their delocalization.
This physical mechanism should be of more general interest and
apply whenever the adsorbed molecule has frontier orbitals that
are symmetry allowed to interact with the atomic orbitals in
atomic wires that participate in the formation of delocalized
states.

\begin{acknowledgments}
Financial support from the Swedish Research Council (VR) and the
Swedish Strategic Foundation (SSF) through the materials
consortium ATOMICS, ISIS, UC Irvine and allocation of computer
resources through SNAC are gratefully acknowledged. The author is
indebted to N. Nilius, W. Ho, D. L. Mills, and R. B. Muniz for
many stimulating discussions and additional insights.
\end{acknowledgments}


\begin{references}
\bibitem{footnoteExcitedState}
In principle, the unoccupied, resonance states of the Au adatom
structures are probed by adding an electron and cannot be directly
described by conventional density functional theory.
\bibitem{cell}
For the CO adsorption on the Au adatom and the ad-trimer, we used
a $3\times 2$ and $5\times 2$ surface unit cell, respectively, and
8 NiAl layers in the slab. To accommodate the adsorbed CO
molecule, we used a vacuum region of about 14.4 {\AA}.The kinetic
energy cut off of the plane wave basis set was about 300 eV and
the surface Brillouin zones for the 3x2 and 5x2 surface unit cells
were sampled uniformly by $6\times6$ and $4 \times 6$ $k$ points,
respectively.
\bibitem{geometry} The C, O, Au atoms and the two
outermost NiAl layers were fully relaxed until the residual forces
were less than 0.05 eV/{\AA}.
\bibitem{PDOS}
The atomic sphere radi were chosen to be 0.73, 0.77, 1.45, 1.25,
and 1.43 {\AA} for the O, C, Au, Ni and Al atoms, respectively.
\bibitem{COgeom}
The C-O bond length is about {1.15 \AA} upon adsorption and the
O-Au bond length is 1.99 and 1.98 {\AA} for structures (A) and
(B), and for structure (C), respectively, in
Fig.~\protect\ref{fig:geom}.
\end{references}
\end{document}